\documentclass[preprint,12pt]{elsarticle}

\usepackage{amssymb}
\usepackage{lipsum}
\usepackage{amsthm}

\usepackage{lineno}
\usepackage{color}

\usepackage{physics}

\journal{VCI2025}

\begin{document}

\begin{frontmatter}



\title{Quantum measurement systems and applications to particle physics and cosmology}%\\\cred{(8 pages in total)} }


\author[first,second]{Masashi Hazumi}
\affiliation[first]{organization={Center for Quantum-field Measurement Systems for Studies of the Universe and Particles (QUP) and Institute of Particle and Nuclear Studies (IPNS), High Energy Accelerator Research Organization (KEK)},%Department and Organization
            addressline={1-1 Oho}, 
            city={Tsukuba},
            postcode={305-0801}, 
            state={Ibaraki},
            country={Japan}}
\affiliation[second]{organization={Department of Physics and Center for High Energy and High Field Physics, National Central University},
            addressline={Zhongli District}, 
            city={Taoyuan City},
            country={Taiwan}}

\begin{abstract}
There are new detector proposals and R\&D that utilize quantum enhancements not previously adopted. Examples include superconducting quantum sensors, atom interferometry, and quantum spin sensors. They are mainly motivated by industrial applications toward quantum computing, secure quantum communication systems, and high-sensitivity sensors. Given the excellent potential of the new quantum measurement systems, there are also new proposals to apply them in particle physics and cosmology. In this review, I survey currently available and emerging quantum technologies and their applications. I then discuss future directions and new proposals for particle physics and cosmology.
\end{abstract}




\end{frontmatter}

%\tableofcontents



%%%%%%%%%%%%%%%%%%%%%%%%%%%%%%%%%%%%%%%%%%%%%%%%%%%%%%%%%%%%%%%%
\section{Introduction}
\label{sec:introduction}

What is quantum sensing?
In quantum metrology,
the term ``quantum sensing"
typically refers to one of the following~\cite{Degen:2016pxo}:
\begin{enumerate}
\item Use of a quantum object to measure a physical quantity (classical or quantum). The quantum object is characterized by quantized energy levels. Specific examples include electronic, magnetic, or vibrational states of superconducting qubits or spin qubits, neutral atoms, or trapped ions.
\item Use of quantum coherence (i.e., wavelike spatial or temporal superposition states) to measure a physical quantity.
\item Use of quantum entanglement to improve the sensitivity or precision of a measurement, beyond what is possible classically.
\end{enumerate}
Here the term ``qubit," or quantum bit, is typically a two-state quantum system with $\ket{0}$ as the lower energy state and $\ket{1}$ as the higher energy state. 
It is a basic element for quantum computers, quantum communication and quantum sensing. 
Quantum sensing exploits changes in the transition frequency or the transition rate in response to an external signal.

Entanglement is one of the most exciting quantum resources
for quantum computing, quantum communication, and quantum sensing.
The following is a conceptual example of how to use entanglement for sensing~\cite{Wineland:1994ngr, Huelga:1997mw}:

A set of $N$ entangled qubits is initialized in a 
Greenberger-Horne-Zeilinger state (GHZ state), defined as
\begin{equation}
\ket{\Psi(t=0)} \equiv \ket{\Psi_0} = \frac{1}{\sqrt{2}}\ket{00\cdots0}+\frac{1}{\sqrt{2}}\ket{11\cdots1}.
\end{equation}
The GHZ state can be in various environments and sense them.
Here we take a typical example, spin qubits in the magnetic field.  
The interaction Hamiltonian, $H$, to describe the system is
\begin{equation}
H \equiv \sum^{N}_{j=1}\frac{\omega}{2}\hat{\sigma}_z^{(j)}.
\end{equation}
Here, $\omega$ is proportional to the magnetic field, which is the quantity to be measured via the GHZ state.
The term $\hat{\sigma}_z^{(j)}$/2 is a spin operator for the $z$ coordinate and for the qubit $j$. 
The time evolution is governed by the Schr\"{o}dinger equation.
At time $t$, the state is
\begin{eqnarray}
\label{eq:evolution}
\ket{\Psi(t)} & = & e^{-iHt}\ket{\Psi_0}\\
 & = & \frac{1}{\sqrt{2}}e^{i\frac{\omega}{2}Nt}\ket{00\cdots0}+\frac{1}{\sqrt{2}}e^{-i\frac{\omega}{2}Nt}\ket{11\cdots1}
\end{eqnarray}
We perform a measurement with a projection, 
$\hat{P}_\perp \equiv \ket{\Psi_\perp}\bra{\Psi_\perp}$, where
we define
\begin{equation}
\ket{\Psi_\perp} \equiv \frac{1}{\sqrt{2}}\ket{00\cdots0}-i\frac{1}{\sqrt{2}}\ket{11\cdots1}.
\end{equation}
The measurement expectation value is obtained as
\begin{eqnarray}
|\bra{\Psi(t)}\hat{P}_\perp\ket{\Psi(t)}|^2 & = & \frac{1}{2} + \frac{1}{2}\sin{N\omega t}\\
& \simeq & \frac{1}{2} + \frac{1}{2}N\omega t,
\end{eqnarray}
where we assume that the time duration $t$ is short enough
that $N\omega t << 1$ holds. 
The remarkable point is that the observed value is enhanced
by a factor $N$ compared with a single spin state case.
This means that the precision of our measurement of the magnetic field is better by a factor $N$.
If we use a set of $N$ separable (i.e. non-entangled, independent) spins for the same magnetic field, the benefit of having $N$ spins over a single spin is a factor $\sqrt{N}$ due to the intrinsic quantum fluctuation of the spin, which is called the spin projection noise.
Therefore, the entanglement gives an improvement of $ N/\sqrt{N}=\sqrt{N}$.\footnote{To be provocative, in the future, humankind might be able to manipulate Avogadro's number of qubits, i.e., $N=6.02\times 10^{23}$. Then the enhancement factor is approximately $10^{12}$.}
As shown in Eq.~(\ref{eq:evolution}), 
the fundamental role of the Schr\"{o}dinger equation is the phase rotation.
For the entangled state, the phase rotation speed is proportional
to the number of qubits, $N$.
Thus, the expected value of the observable is also proportional to $N$.

Note that the argument above ignores practical limitations.
In particular, entanglement is fragile against decoherence. However,
entanglement-enhanced magnetometry is still possible under decoherence, which can pave a way for new applications in particle physics and cosmology~\cite{Sichanugrist:2024wfk}.
Sensors using entanglement are not realized today due to technological challenges.
But the example above clearly demonstrates the huge potential of entanglement and qubits
for future sensing.

Particle physicists including myself tend to think that
all the detector technologies we have been using in particle physics
are ``quantum sensing" as they
need quantum physics to describe the operating principles.
As shown above, the new quantum sensing (part of ``Quantum 2.0" as described in
the next section) today is qualitatively different,
pursuing a new level of quantum physics applications such as the extensive use of entanglement, which 
have never been done in particle physics before.
For many detector technologies we have in hand in particle physics today, classical physics intuition can be applied, e.g., to understand behaviors of electrons, photons, and quantum physics is often playing a supporting role to provide us with precise calculations.
In the new quantum sensing, quantum physics rules the operating principles beyond the classical physics intuition.
The potential gain of new quantum methodologies is enormous, which may lead to fundamental innovation.

In addition to the spin qubit mentioned above, other interesting principles for qubits have been proposed, including superconducting qubits, trapped ion qubits,
topological qubits, photonic qubits, and qubits with neutral atoms. In any case,
qubits must be initialized, read out, coherently manipulated, and interact with a relevant physical quantity~\cite{Degen:2016pxo}.

The definition of quantum sensing described above is from the quantum metrology community,
which might be too narrow when we consider all the possible applications
to particle physics, high energy physics (HEP) and cosmology.
An alternative definition of quantum sensing techniques and sensors for HEP is found, for example, in~\cite{Chou:2023hcc}:
``New and emerging quantum sensing technologies and methodologies which have not been traditionally used in HEP experiments and whose development could efficiently leverage large investments being made worldwide in quantum science."
With this definition, examples of quantum sensors include the following:
\begin{itemize}
\item Atom interferometry; 
\item Atomic, nuclear, and molecular clocks and optical cavities; 
\item Superconducting nanowire single-photon detectors (SNSPDs); 
\item Superconducting qubits; 
\item Continuous variables quantum sensors and amplifiers; 
\item Superconducting cavities; 
\item Qubit-based pair-breaking detectors; 
\item Quantum capacitance detectors; 
\item Superconducting quasiparticle-amplifying transmon; 
\item Kinetic inductance detectors (KIDs); 
\item Transition edge sensors (TES); 
\item Spin sensors and NMR; 
\item Superfluid helium sensors; 
\item Optomechanics; 
\item Quantum networks and long-distance quantum coherence; 
\item Optical interferometry and precision astrometry; 
\item Dark matter detection and quantum networked sensors.
\end{itemize}
Following the above definition, there are many types of quantum sensors, and readers can even invent a new quantum sensor idea and add it to the list.

In this document, I use the term ``quantum measurement system" as a system that contains quantum sensing techniques and quantum sensors as an essential component.
The system shall satisfy a project's requirements flow, from science to instruments to sub-components, so that it can do its job in realistic, sometimes harsh, environments.
Proper system considerations are essential for practical applications.
It is often the case that the subsystems, such as the readout subsystem,
mechanical subsystem, become a bottleneck of the development, even if quantum sensors as a component are already established technologically.
It is sometimes necessary to devise creative solutions to overcome various challenges.
We should never underestimate the importance of system development
when we discuss the applications of the new quantum sensors.

Communities of particle physics, astrophysics/cosmology, and quantum information science (QIS) have overlaps, but each community has its own agenda and culture. Understanding the culture of neighboring communities is particularly important for fruitful joint studies.
In the next section, we provide an overview of the major activities
of the QIS community.

%%%%%%%%%%%%%%%%%%%%%%%%%%%%%%%%%%%%%%%%%%%%%%%%%%%%%%%%%%%%%%%%
\section{Quantum Information Science (QIS) overview}
\label{sec:qis}

\begin{figure}[ht]
	\centering 
	\includegraphics[width=\textwidth]{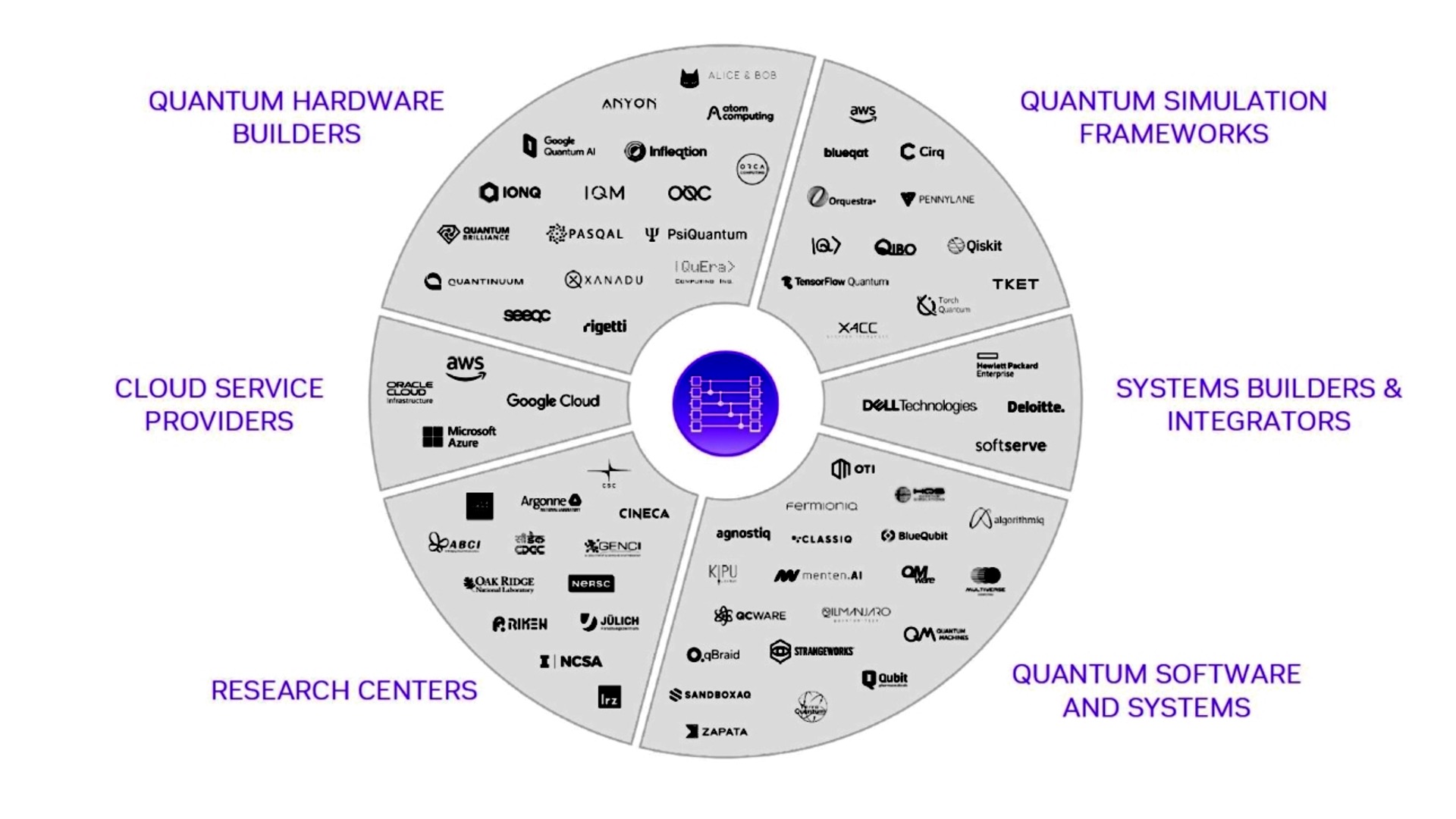}	
%MH \includegraphics[width=0.48\textwidth]{quantumecosystem.jpg}	
%    \rule{8cm}{6cm}
	\caption{Quantum Ecosystem concept~\cite{nvidia}.} 
	\label{fig:quantumecosystem}%
\end{figure}

There are four areas of activities in QIS today\footnote{Quantum Information Science and Technology (QIST) is another term often used in the literature.}: 
Quantum computing;
Quantum communications (``quantum internet");
Quantum sensing;
Quantum materials.
%Quantum Information Science (QIS) is an ever-evolving field. 
%It should be noted that this proceedings may also be out of date. 
%Please be aware that this is only a snapshot as of February 2025.
%
A new conference series ``Quantum 2.0" has been launched recently.
The following is a foundational statement about the conference~\cite{optica:quantum}:
``Quantum 2.0 refers to the development and use of many-particle quantum superposition and entanglement in large engineered systems to advance science and technology. Examples of such large quantum systems are quantum computers and simulators, quantum communication networks, and arrays of quantum sensors. New resulting technologies will go far beyond the (quantum 1.0) capabilities." 
Those who are involved in Quantum 2.0 are keen on
establishing the ``Quantum Ecosystem"
(Fig.~\ref{fig:quantumecosystem}~\cite{nvidia} as an example).
The idea here is that the ecosystem consists of
quantum hardware builders, cloud service providers, research centers,
quantum simulation frameworks, system builders \& integrators,
and quantum software and systems to establish sustainable
development of quantum technology.
The global quantum effort leading to research and innovation in quantum science and technology is continually rising with current worldwide investments exceeding \$40 billion. Overall, the global quantum technology market is projected to reach \$106 billion by 2040~\cite{qureca}.
In the following, I show four examples of recent achievements in QIS.

\begin{figure}[ht]
	\centering 
	\includegraphics[width=\textwidth]{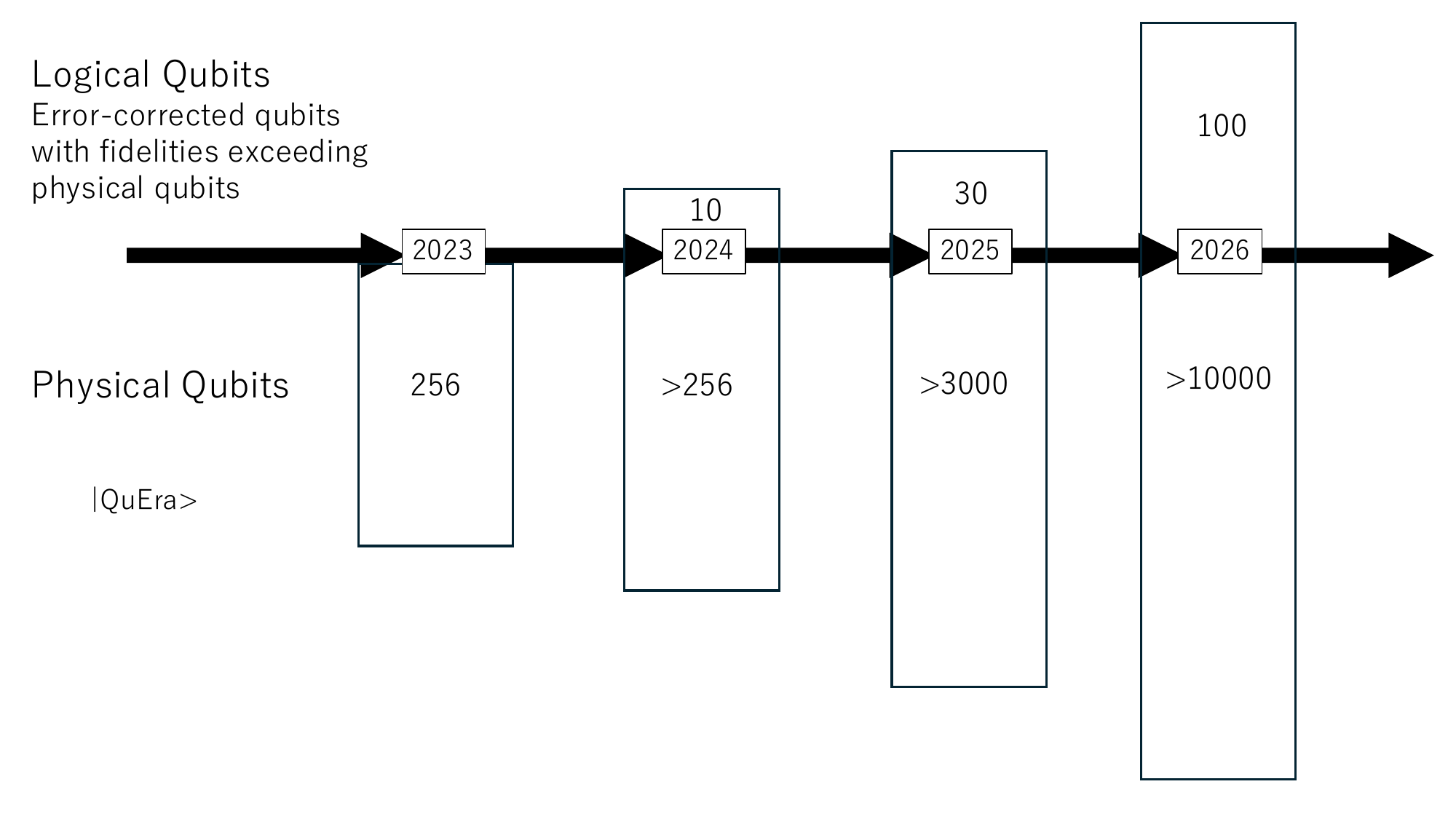}	
%MH	\includegraphics[width=0.48\textwidth]{roadmap.pdf}	    
%    \rule{8cm}{6cm}
	\caption{Quantum computing roadmap. Information is taken from~\cite{QuEraRoadmap2024}.} 
	\label{fig:qubitloadmap}%
\end{figure}

The first example is an advancement in quantum computation.
The Physics World Breakthrough of the Year was awarded,
from all the areas of physics, to two advances in
quantum error corrections for quantum computing~\cite{PhysicsWorldBreakThrough2024}.
One is to researchers at Harvard, MIT and QuEra Computing
``for demonstrating quantum error correction on an atomic processor with 48 logical qubits,"~\cite{Bluvstein:2023zmt} and the other to
researchers at Google Quantum AI
``for demonstrating quantum error correction below the surface code threshold in a superconducting chip"~\cite{GoogleQuantumAIandCollaborators:2024efv}.
The quantum computing community is now more keen on quantum error correction to get enough logical qubits, rather than increasing the number of physical qubits. Figure~\ref{fig:qubitloadmap}
shows an example of a quantum computing roadmap by QuEra~\cite{QuEraRoadmap2024} that shows the target number of logical qubits in addition to the number of physical qubits.

The second example is an interesting achievement in 2024 on quantum communication, the first demonstration of quantum teleportation over conventional fiber-optic cables carrying internet traffic~\cite{Thomas:2024yhq}.
The authors demonstrated quantum state teleportation over a 30.2-km fiber that was populated with high-power 400-Gbps conventional data traffic, a significant step towards ensuring that complex multi-photon/multi-node quantum network applications could be realized anywhere in the existing fiber infrastructure.
Marconi's first radio broadcast demonstration more than 100 years ago paved the way for the radio and television broadcasts that we take for granted today.
The work above might be an equivalent event of the century.
Only people in the 22nd century will know if this is the case.
%People in 22nd century will tell if this is the case.

The third example is the development of quantum sensing for quantum communication.
The progress of superconducting nanowire single-photon detectors (SNSPDs) has been remarkable in recent years. 
Significant improvements are seen on
timing jitter, intrinsic photon number resolution,
efficiency, array size, maximum count rate, dark count rate, active area,
and cut-off wavelength~\cite{Korzh:2018oqv, Reddy:2020usi, Oripov:2023yod, Craiciu:2022akn, Chiles:2021gxk, Taylor2023},
already at a level of significant interest in
%already at the level very interesting in 
particle physics applications. 
Tests with a 120 GeV proton beam have been carried out~\cite{Lee:2023brm}.

\begin{figure}[ht]
	\centering 
	\includegraphics[width=\textwidth]{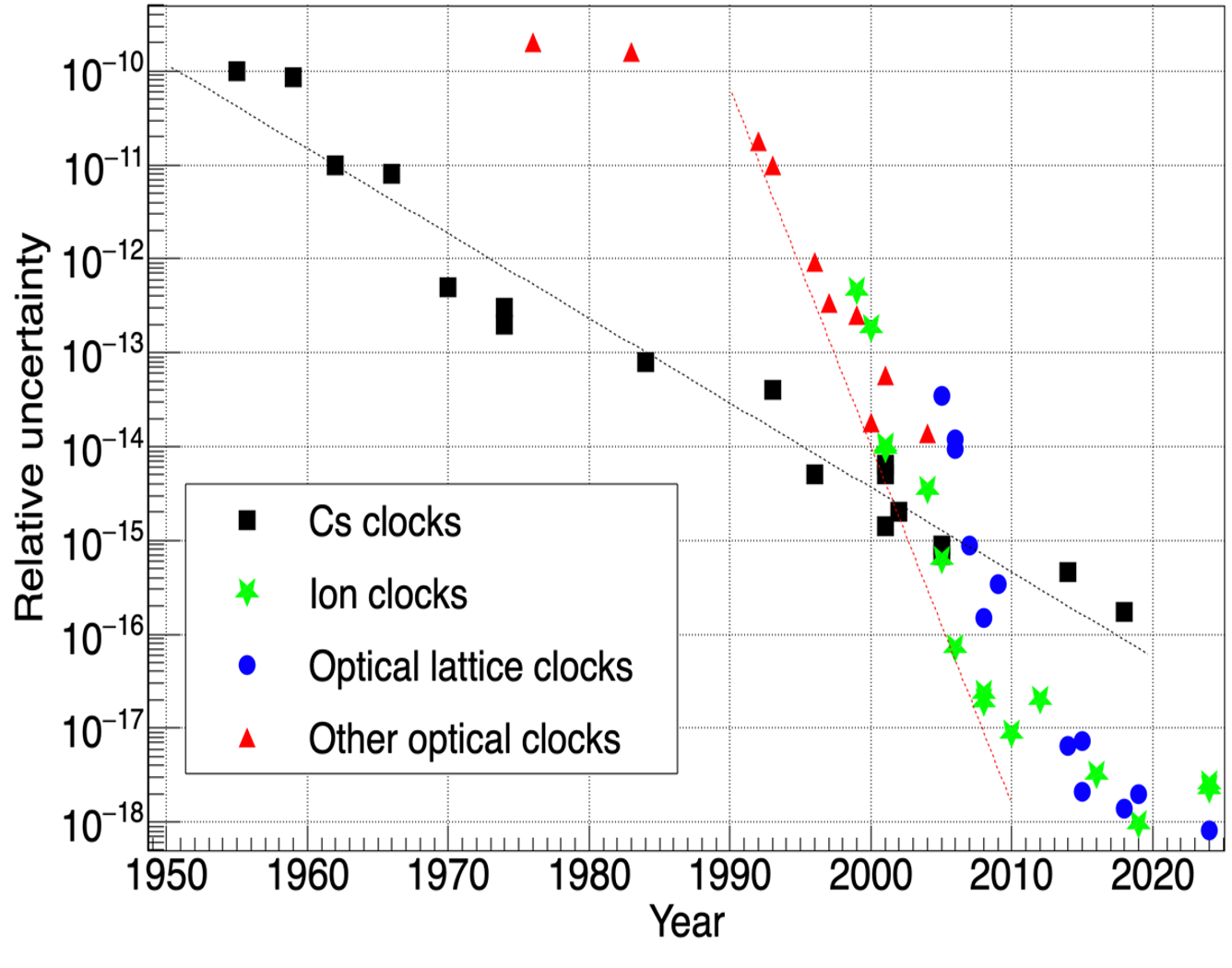}	
%MH	\includegraphics[width=0.48\textwidth]{clock.png}	
%    \rule{8cm}{6cm}
	\caption{Clock sensitivities over time~\cite{kawasaki2025quantumsensingusingcold}. The black dotted line shows the average improvement in the uncertainty for the Cs clocks, and the red dotted line shows approximate development of the uncertainty for the best optical clocks.
} 
	\label{fig:clock}%
\end{figure}
The fourth example is the development of an atomic clock
with $8\times 10^{-19}$ systematic uncertainty~\cite{PhysRevLett.133.023401}.
Figure~\ref{fig:clock} shows 
the achieved relative sensitivities of atomic clocks over time~\cite{kawasaki2025quantumsensingusingcold}.
This level of accuracy allows the clock to detect even minute disturbances,
%The clock with this accuracy is sensitive to any disturbance, 
including that from new physics.
The accuracy is thus already interesting enough to search for new physics.

Quantum sensing for biomedical applications is another active 
research area. 
%Topics included are 
Topics include the determination of molecular structures, thermal measurements using nanodiamonds, 
%molecular structure determination, thermal measurements with nanodiamonds, 
measurements of in vivo magnetic activity in animals,
subcellular organelle metabolic studies, electrical activity studies
in cellular cultures, and clinical diagnostics in humans~\cite{Aslam:2023woi}.

%%%%%%%%%%%%%%%%%%%%%%%%%%%%%%%%%%%%%%%%%%%%%%%%%%%%%%%%%%%%%%%%
\section{Quantum sensors in use in particle physics and cosmology today}
\label{sec:inuse}
Four examples are shown in this section to illustrate 
how quantum sensors are used in particle physics and cosmology today.

\begin{figure}[ht]
	\centering 
	\includegraphics[width=\textwidth]{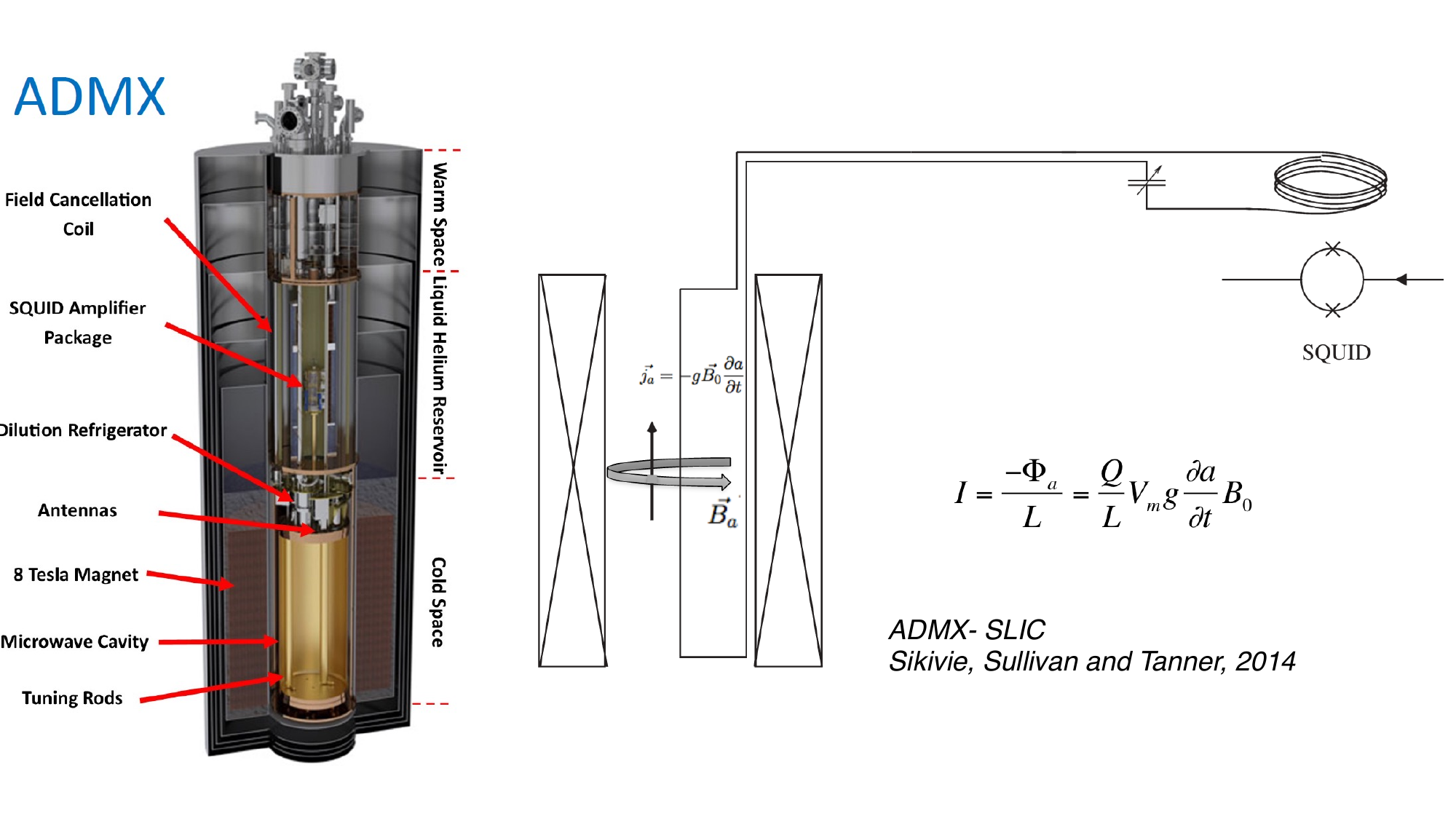}	
%MH	\includegraphics[width=0.48\textwidth]{ADMX.pdf}	
%    \rule{8cm}{6cm}
	\caption{ADMX haloscope and its circuit diagram~\cite{AndrewSonnenschein:quposium2024}.} 
	\label{fig:admx}%
\end{figure}

The first example is the QCD axion search.
Quantum measurement system components such as SQUIDs, parametric amplifiers, and resonant amplification are in use in particle physics.
The ADMX experiment is a typical example, which already touched on the QCD axion predictions in the mass range of $m_a = O(10^{-6})$~eV~\cite{ADMX:2024xbv}.
The axion field $a$ follows axion electrodynamics expressed by the following set of equations:
\begin{eqnarray}
\nabla\cdot E & = & g_{a\gamma\gamma}B\cdot\nabla a,\\
\nabla\times E & = & -\frac{\partial B}{\partial t},\\
\nabla\times B - \frac{\partial E}{\partial t} & = & g_{a\gamma\gamma}\bigg(E\times\nabla a - B\frac{\partial a}{\partial t}\bigg),\\
\nabla\cdot B & = & 0.
\end{eqnarray}
Under these equations, 
the current generated by the axion field is
\begin{equation}
I = -\frac{\Phi_a}{L}=\frac{Q}{L}V_mg_{a\gamma\gamma}\frac{\partial a}{\partial t}B_0,
\end{equation}
(see also Fig.~\ref{fig:admx}).

\begin{figure}[ht]
	\centering 
%MH	\includegraphics[width=0.48\textwidth]{POLARBEAR.pdf}	
	\includegraphics[width=\textwidth]{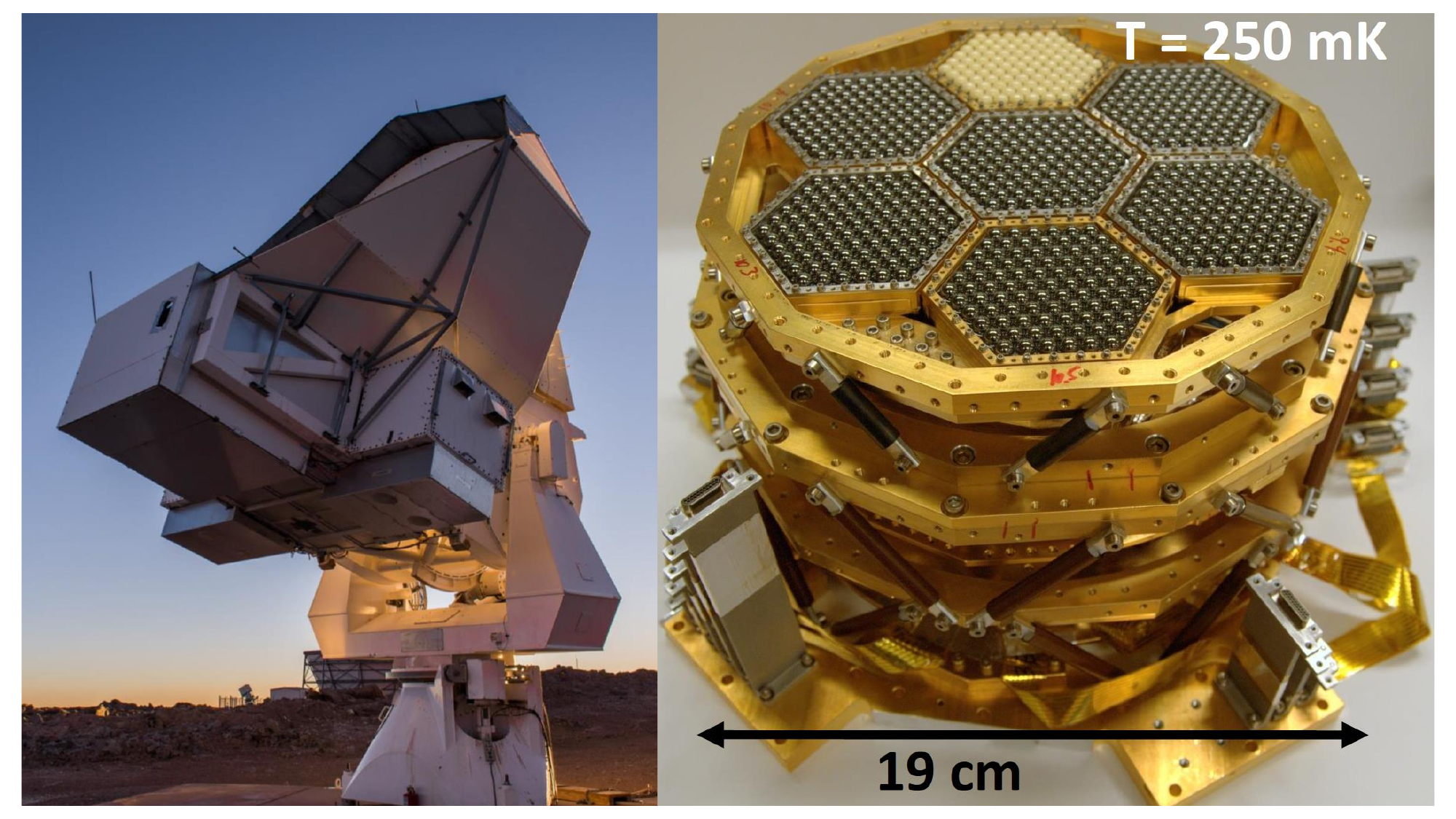}	
%    \rule{8cm}{6cm}
	\caption{The POLARBEAR telescope and the TES bolometer array. The physical diameter of the primary mirror is about 4~m.} 
	\label{fig:polarbear}%
\end{figure}

\begin{figure}[ht]
	\centering 
%MH	\includegraphics[width=0.48\textwidth]{dfmux.pdf}	
	\includegraphics[width=\textwidth]{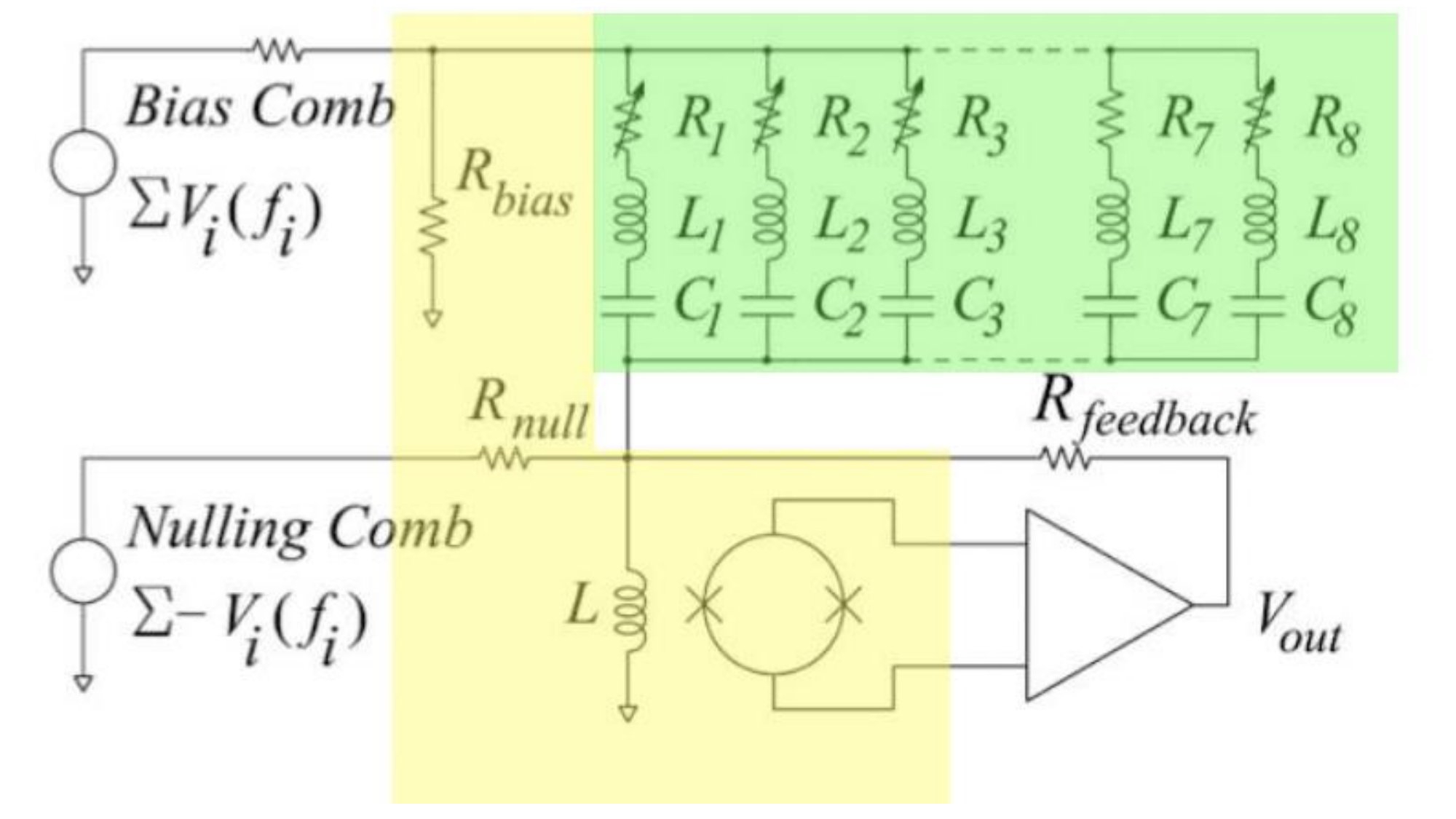}	
%    \rule{8cm}{6cm}
	\caption{Schematic of the frequency multiplexed readout scheme at POLARBEAR.} 
	\label{fig:dfmux}%
\end{figure}

The second example is measurements of the cosmic microwave background (CMB) polarization.
Many projects are ongoing or in preparation.
In Atacama, Chile, POLARBEAR~\cite{POLARBEAR:2014hgp}, Simons Array~\cite{POLARBEAR:2015ixw}, ACTPol~\cite{Thornton:2016wjq},
Simons Observatory~\cite{SimonsObservatory:2018koc, Galitzki:2018wvp}, and CLASS~\cite{Essinger-Hileman:2014pja}, are analyzing data or taking data.
At the South Pole, SPTPol~\cite{Austermann:2012ga}, SPT-3G~\cite{SPT-3G:2021vps}, BICEP/Keck~\cite{Hui:2018cvg, BICEP2:2015nss, BicepKeck:2021ybl} are analyzing data or taking data.
There are also QUIJOTE~\cite{Genova-Santos:2015uia} and GroundBIRD~\cite{GroundBIRD:2020wax} in Teine (Canary Islands),
QUBIC~\cite{QUBIC:2020kvy} in Alto Chorillo (Argentina), and AliCPT~\cite{Ghosh:2022mje} in Tibet (China).
There are also observations with balloons (SPIDER~\cite{SPIDER:2021ncy}, EBEX~\cite{EBEX:2018zpz}, LSPE~\cite{LSPE:2020uos}, and PIPER~\cite{PIPER:2016szc}).
These projects will produce fruitful results in next ten years.
Two large-scale projects are proposed for observations in 2030s.
One is LiteBIRD~\cite{Hazumi2008AIP, Hazumi:2011zz, Hazumi:2012gjy, Matsumura2014JLTP, Matsumura2014SPIE, LiteBIRD:2022cnt}, a JAXA-led satellite project, and
the other is CMB-S4~\cite{Abazajian:2019eic}, a US-led ground telescope array.
One of the most exciting scientific objectives of CMB polarization measurements
is testing cosmic inflation and quantum gravity.
Primordial gravitational waves, generated during cosmic inflation, should have imprinted the curl pattern in the CMB polarization map, called the CMB B-mode, as fingerprints of inflation~\cite{seljak/zaldarriaga:1997,kamionkowski/kosowsky/stebbins:1997,HuWhite1997Total}.
The CMB B-mode is by far the most sensitive probe of the primordial gravitational waves.
Since inflation is a phenomenology, we want to find out particle physics behind it.
The simplest class of models introduce a new inflationary scalar particle called the inflaton, which is expected to have a GUT-scale potential energy.
Therefore, this observation is connected to the GUT-scale physics.
%The $r$ predictions change in the form of potentials, but LiteBIRD achieves a sensitivity that allows us to check a whole set of representative inflation models.
Cosmic inflation might also be connected to the Higgs boson,
as the inflaton is also a spin zero particle.
Models have been proposed assuming that inflation is driven by the Higgs boson~\cite{Bezrukov:2007ep,Ballesteros:2016xej}. 
Future projects such as LiteBIRD can test this hypothesis~\cite{LiteBIRD:2022cnt}.
Figure~\ref{fig:polarbear} shows the POLARBEAR telescope and its TES bolometer array as a typical CMB observation instrument.
POLARBEAR is a dedicated CMB polarization experiment
located at 5200 meters in Atacama, Chile.
It received the first light in Jan. 2012, and was expanded to Simons Array from 2018. 
Figure~\ref{fig:dfmux} shows the circuit diagram of the POLARBEAR detector system. TESes are shown as variable registers $R_1\cdots R_8$.

The third example is the NIKA2 camera~\cite{Adam:2017gba}, which adopts microwave KIDs (MKIDs) for astrophysical observations.
\begin{figure}[ht]
	\centering 
	\includegraphics[width=\textwidth]{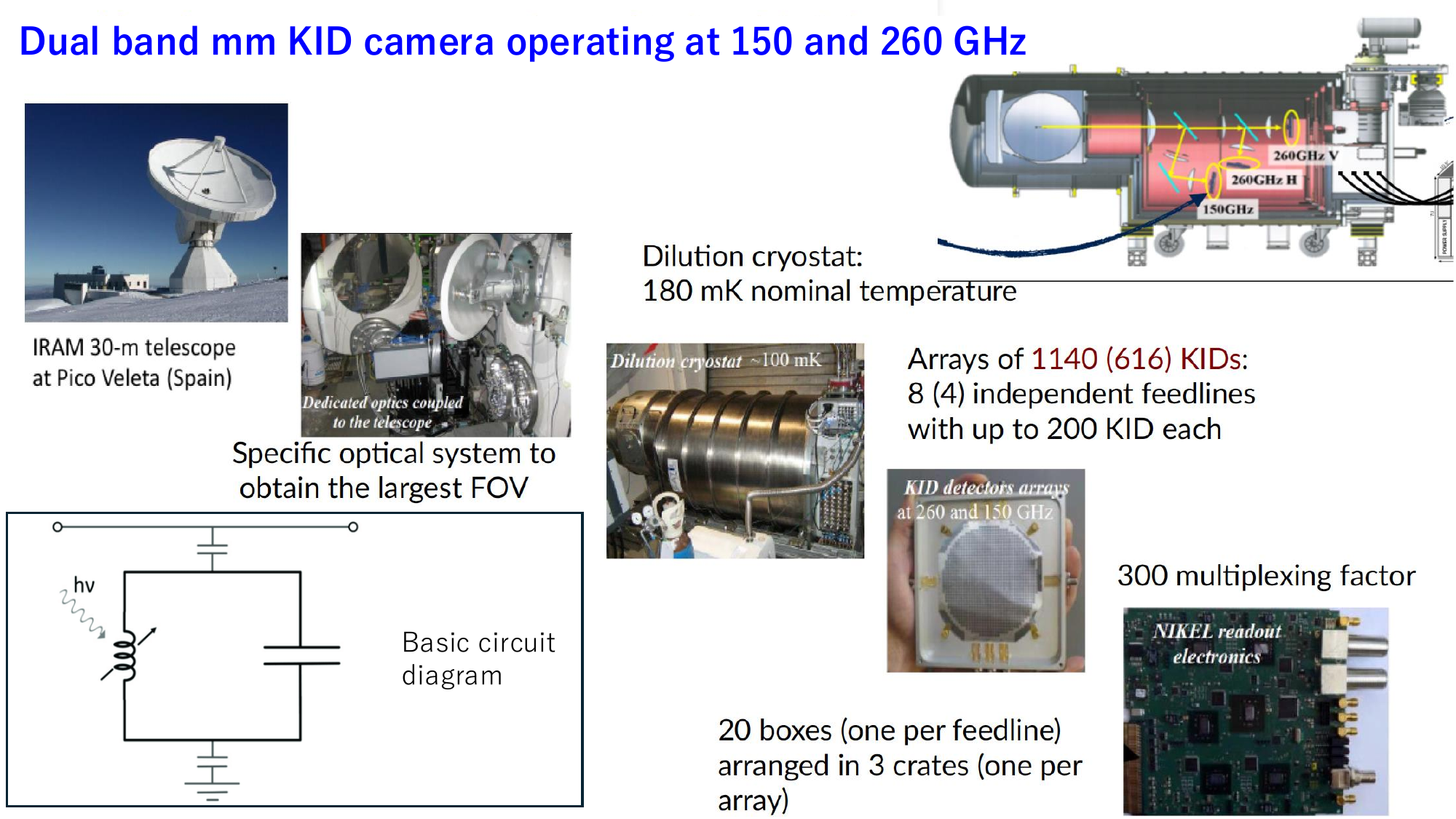}	
%MH	\includegraphics[width=0.48\textwidth]{NIKA2.pdf}	
%    \rule{8cm}{6cm}
	\caption{NIKA2 instrument~\cite{Adam:2017gba} and basic circuit diagram of MKID.} 
	\label{fig:nika2}%
\end{figure}
Figure~\ref{fig:nika2} shows the NIKA2 instrument and its basic MKID circuit diagram. The variable inductance shown in the inset of the figure corresponds to an MKID.
When incoming photons break Cooper pairs in the MKID,
the kinetic inductance changes. With a proper LC circuitry,
the change can be detected as a change of the phase or amplitude of
the output signal. This scheme allows us to multiplex a large number of detectors, which is crucial to keep the heat inflow from
the room-temperature stage to the cryogenic stage low enough.

The fourth example is a TES array on the beamline for nuclear physics.
A feasibility test measurement with a low-density neon gas target was performed by observing X-rays emitted by muonic neon via the transition, 6.3 keV, using a multi-pixel array TES microcalorimeters at the J-PARC muon facility~\cite{Okada:2020fjo}.

%%%%%%%%%%%%%%%%%%%%%%%%%%%%%%%%%%%%%%%%%%%%%%%%%%%%%%%%%%%%%%%%
\section{Future prospects}
\label{sec:future}

Multiple workshops have been held recently on 
quantum sensing for HEP.
Examples include the following:
\begin{itemize}
\item Quantum Sensors for HEP (27–29 Apr. 2023)~\cite{Chou:2023hcc};
\item QUPosium2024 at KEK (11-13 Dec. 2024)~\cite{QUPosium2024};
\item QT4HEP2025 at CERN (20-24 Jan. 2025)~\cite{QT4HEP2025};
\item DRD5/RDquantum collaboration meeting at CERN (17-19 Feb. 2025)~\cite{DRD5meeting}.
\end{itemize}
Science drivers in fundamental physics for quantum sensors
at the ``Quantum Sensors for HEP" workshop were the following:
\begin{itemize}
\item Dark waves;
\item Dark particles;
\item Cosmology, dark energy, phase transitions
(targeting improved sky observations for HEP science,
direct detection of dark energy, and
inflation and gravitational waves);
\item Quantum gravity;
\item Testing quantum mechanics;
\item Beyond the standard model (BSM) physics with accelerators and colliders;
\item Fundamental symmetries and interactions.
\end{itemize}
%Dark waves; Dark particles; Cosmology, dark energy, phase transitions; Quantum gravity; Testing quantum mechanics; Beyond the standard model (BSM) physics with accelerators and colliders; Fundamental symmetries and interactions.

Key elements of the concordance model in cosmology ($\Lambda$CDM)
are missing in the standard model of particle physics (the SM).
That is the biggest issue in the quest for the ultimate theory of nature
and source of our enthusiasm today.
To be more specific, we summarize the current situation as follows:
\begin{itemize}
\item Five mysteries require physics beyond the SM;
    \begin{itemize}
    \item Cosmic inflation (accelerating expansion in the primordial universe);
    \item Baryon asymmetry (baryogenesis? leptogenesis?);
    \item Neutrino properties (mass?, Dirac vs. Majorana, $N_{eff}$, sterile?);
    \item Dark matter (heavy or light? SUSY or else?);
    \item Dark energy (accelerating expansion in the late universe).
    \end{itemize}
\item There are tantalizing hints of beyond $\Lambda$CDM anomalies;
    \begin{itemize}
    \item Hubble tension~\cite{DiValentino:2021izs};
    \item Deviation from the cosmological constant~\cite{DESI:2024mwx}.
    \end{itemize}
\end{itemize}
A resolution of any of the mysteries and anomalies above will revolutionize our picture of the Universe. New quantum measurement systems are needed to make an observational breakthrough.

Some R\&D programs are running to enhance connections between QIS, HEP and cosmology, including those led by CERN, NASA, and US-HEP.
Small-scale and/or individual/inventive R\&Ds/projects are also active.
In the following, I show some examples to illustrate what could happen in the future.

DRD5 (Detector R\&D Collaboration)~\cite{DRD5} 
is part of a large effort to develop detectors for future particle physics experiments in a number of technologies (thus DRD1$\cdots$DRD8).
DRD5 focuses on R\&D of Quantum Sensors.
The DRD5 proposal was approved in 2024 by CERN's Research Board, after having been recommended by CERN's DRDC (Detector R \& D Committee). DRD5 has the following 6 work packages to
ensure that all sensor families that were
identified in the ECFA roadmap as relevant to future advances
in particle physics are included:
WP1: Atomic, nuclear and molecular systems in traps \& beams;
WP2: Quantum materials (0-, 1-, 2-D);
WP3: Quantum superconducting devices;
WP4: Scaled-up massive ensembles (spin-sensitive devices, hybrid devices, mechanical sensors;
WP5: Quantum techniques for sensing;
WP6: Capacity expansion.

\begin{figure}[ht]
	\centering 
	\includegraphics[width=\textwidth]{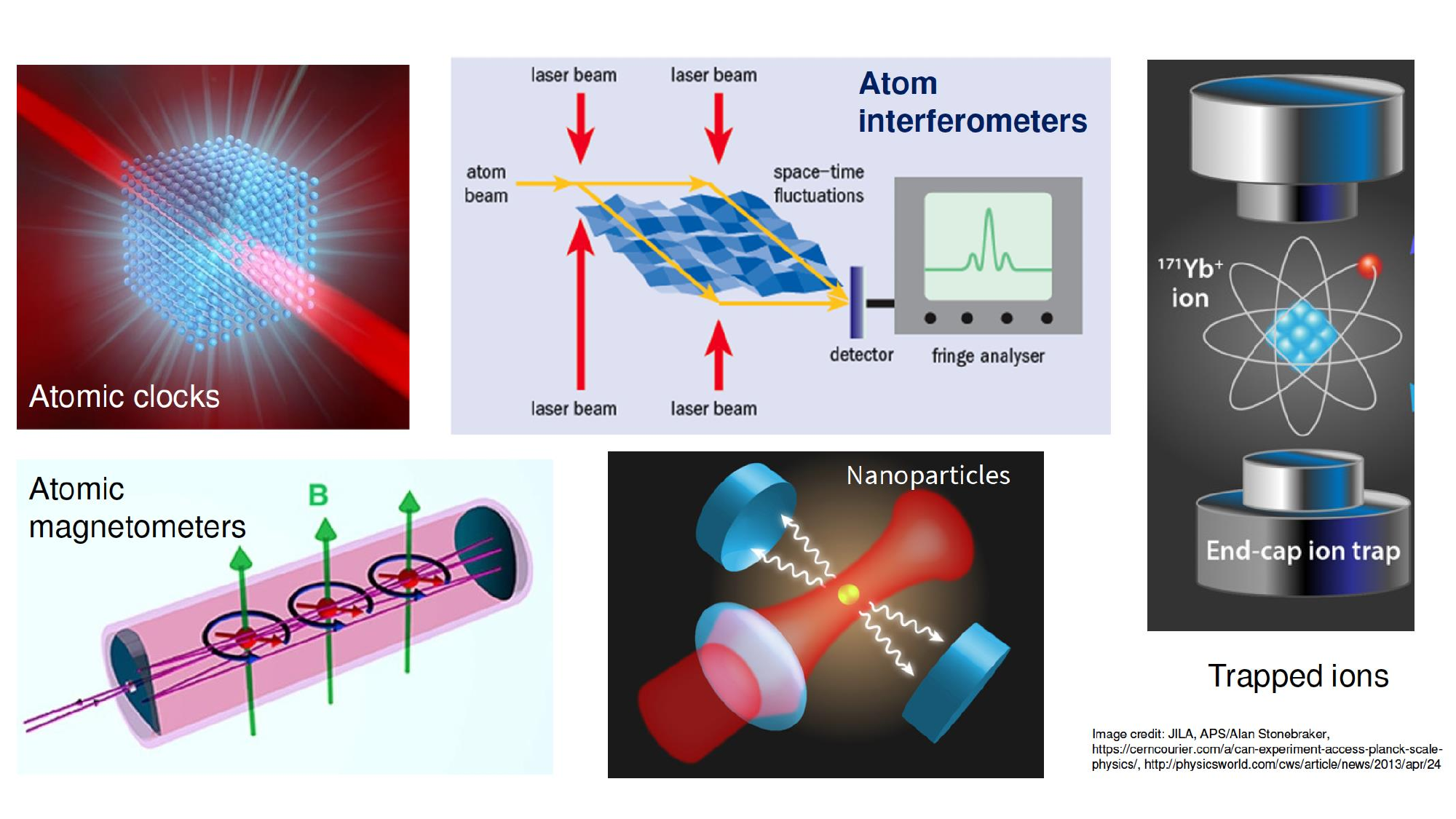}	
%MH	\includegraphics[width=0.48\textwidth]{atom.pdf}	
%    \rule{8cm}{6cm}
	\caption{Atom-based quantum sensors.} 
	\label{fig:atombasedquantumsensor}%
\end{figure}

R\&D on atom-based quantum sensors, including
atomic clocks, atom interferometers, atomic magnetometers,
sensors with trapped ions, is a very active area
(Fig.~\ref{fig:atombasedquantumsensor}).
Atom interferometry is a new emerging technology.
It is conceptually similar to interferometry for photons,
with beam splitters and mirrors for atoms,
and is potentially sensitive to gravitational waves
and dark-matter waves.
While the ultimate goal is to construct 
a facility with an O(km) size,
the current R\&D focuses on demonstrating O(10)-O(100)m prototypes,
such as AION~\cite{Badurina:2019hst}
and MAGIS~\cite{MAGIS-100:2021etm}.

\begin{figure}[ht]
	\centering 
	\includegraphics[width=\textwidth]{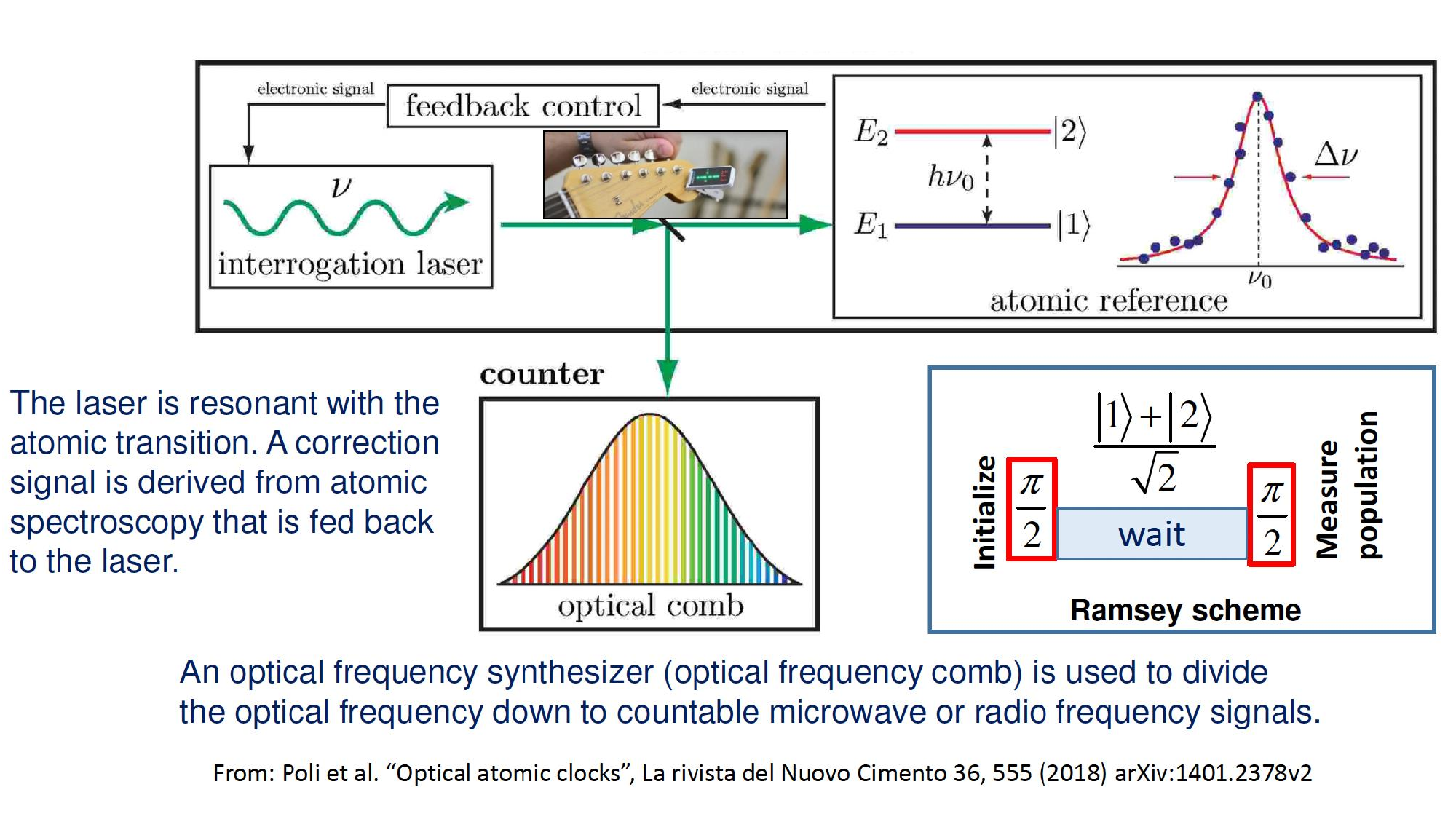}	
%MH	\includegraphics[width=0.48\textwidth]{atomclock.pdf}	
%    \rule{8cm}{6cm}
	\caption{Operating principle of atomic clocks.} 
	\label{fig:atomclock}%
\end{figure}

Atomic clocks (and future nuclear clocks) are also promising
tools to test fundamental physics.
Gravity, velocity, and unseen forces can affect the tick rate of atomic clocks.
By comparing their tick rate with that of other clocks, ultra-precise atomic clocks can search for changes and deviations in gravity, new forces, and variations to fundamental constants.
R\&D on nuclear clocks is also ongoing, 
whose accuracy could, in principle, be better by approximately three orders of magnitude than that of atomic clocks.
Figure.~\ref{fig:atomclock} illustrates the operating principle of the atomic clock.

NASA develops a broad range of technologies to support space-based quantum sensing and communications, uses the space environment to study fundamental quantum processes to advance our knowledge of physics~\cite{Mercer:2025hfx}.
With multiple satellites, 
quantum sensors can be deployed in an entangled state on a gigantic scale.
Future use of networked-entangled sensors may lead to measurements that are not currently possible, leading to dramatic improvements in our ability to understand the Universe.

%%%%%%%%%%%%%%%%%%%%%%%%%%%%%%%%%%%%%%%%%%%%%%%%%%%%%%%%%%%%%%%%
\section{An example of serendipity}
\label{sec:serendipity}

Although community-wide planning described in the previous section is necessary, serendipity is also essential.
Historically, there are many examples of unexpected discoveries paving the way to new directions. 
This section describes a recent example of unexpected results the author encountered, and a related new idea to search for ultra-light axion-like particles (ALPs).

The Crab Nebula, also known as Tau A in radio astronomy, is a polarized astronomical source at millimeter wavelengths.
The POLARBEAR observed Tau A periodically as a calibration tool
for the polarization angle.
We found a hint of oscillation at 2.5$\sigma$ (including look-elsewhere-effect) with a frequency of around 2 months~\cite{POLARBEAR:2024vel}. 
If this is due to the ALP-photon coupling, the corresponding ALP mass is around $m_a \sim 10^{-21}$~eV.

\begin{figure}[ht]
	\centering 
	\includegraphics[width=\textwidth]{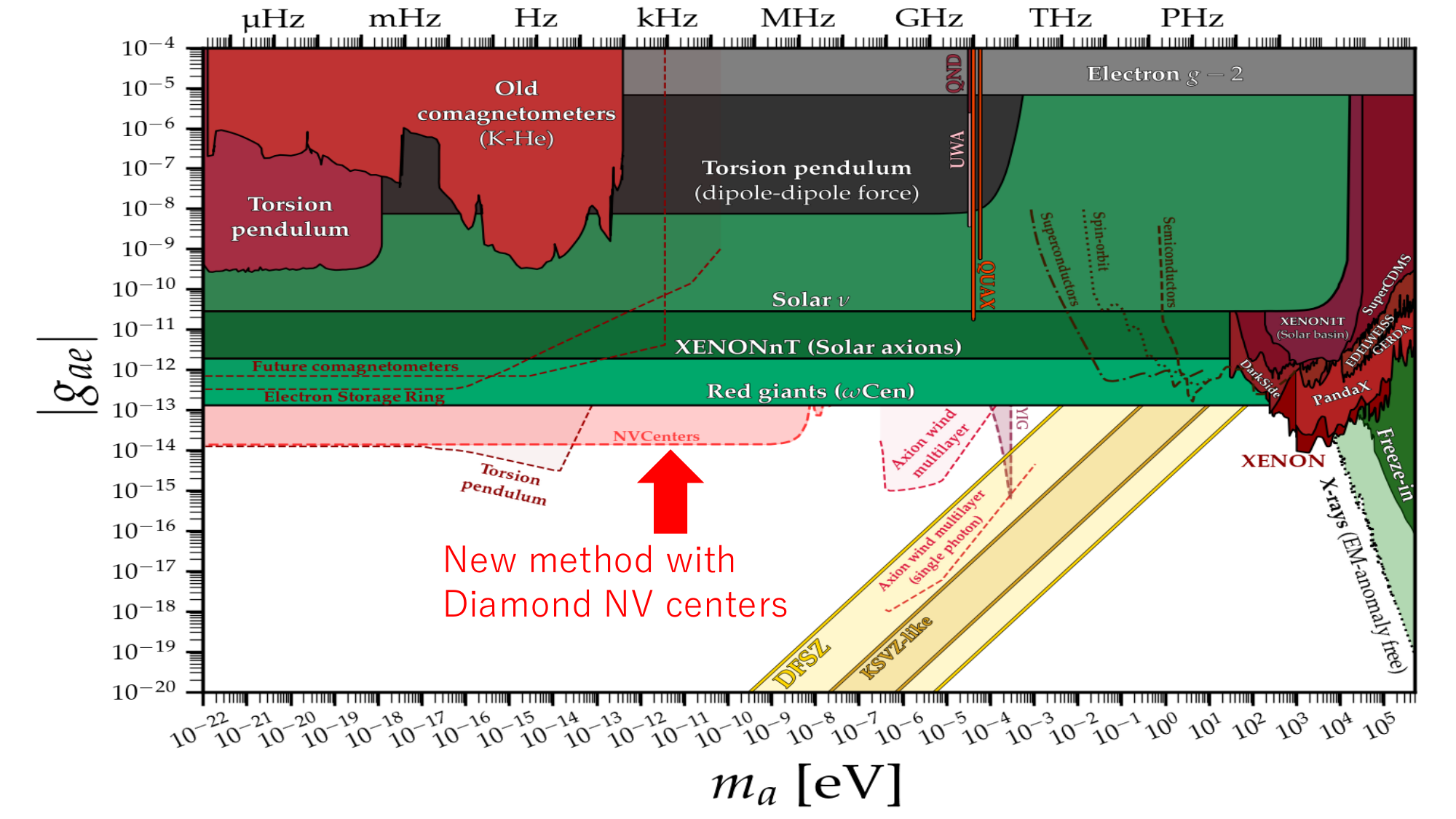}	
%MH	\includegraphics[width=0.5\textwidth]{alpelectroncoupling.pdf}	
%    \rule{8cm}{6cm}
	\caption{Current limits and future prospects on axion-electron coupling measurements~\cite{axionlimits}.} 
	\label{fig:axionelectroncoupling}%
\end{figure}

In parallel, the author formed an interdisciplinary 
team of scientists in particle physics and QIS communities
to propose a new method to search for ALPs by
using spin qubits with diamond NV centers~\cite{Chigusa:2023roq, Chigusa:2024psk, Herbschleb:2024pbk}.
As shown in Fig.~\ref{fig:axionelectroncoupling},
the new method is sensitive to the ALP-electron coupling,
which is complementary to the method at POLARBEAR.
ALP fields are similar to magnetic fields, as
the effective Hamiltonian is
\begin{equation}
H_{\rm eff} = \frac{g_{aee}}{m_e}\nabla a\cdot \vec{S}_e.
\end{equation}
Future measurements with scalable NV diamond quantum sensors are promising. Experimental attempts to carry out initial measurements have started recently.

%%%%%%%%%%%%%%%%%%%%%%%%%%%%%%%%%%%%%%%%%%%%%%%%%%%%%%%%%%%%%%%%
\section{Summary}
\label{sec:summary}

There are new proposals and R\&D projects that leverage previously unused quantum enhancements. Examples include superconducting quantum sensors, atom interferometry, and quantum spin sensors. They are mainly motivated by industrial applications toward secure quantum communications systems, quantum computing, and high-sensitivity sensors. 
These advances have also led to novel proposals for applications in particle physics and cosmology.
%Given the excellent potential of the new quantum measurement systems, there are also new proposals to apply them in particle physics and cosmology. 
In this review, I have surveyed currently available and emerging technologies and their applications for particle physics and cosmology.
The following are the main messages.
\begin{itemize}
\item The use of quantum objects, characterized by quantized energy levels with the capability of initialization, readout, and manipulation, has huge potential to measure physical quantities, including those in particle physics and cosmology.
\item The QIS community is growing. New technologies are coming rapidly.
\item Some fundamental physics drivers, e.g., ultra-light dark matter candidates, including axion-like particles, are very suitable for quantum sensing.
\item You’re never too late to jump into R\&D on quantum sensing for particle physics and cosmology.
The best way to predict the future is to invent it.
\end{itemize}

\section*{Acknowledgements}
I thank Prof. Yuichiro Matsuzaki for checking the draft and providing valuable comments.
I am grateful to the organizers of VCI2025.
My participation to VCI2025 was supported by JSPS KAKENHI Grant No.
22H04945.

%\appendix
%
%%%%%%%%%%%%%%%%%%%%%%%%%%%%%%%%%%%%%%%%%%%%%%%%%%%%%%%%%%%%%%%%
%\section{Appendix 1}
%\label{app:appone}
%\bibliographystyle{elsarticle-harv} 
%\bibliographystyle{plain}
%\bibliographystyle{apsrev4-1}
\bibliographystyle{elsarticle-num} 
\bibliography{references}



%% For example:

%% \bibitem[Aladro et al.(2015)]{Aladro15} Aladro, R., Martín, S., Riquelme, D., et al. 2015, \aas, 579, A101

%%\end{thebibliography}

\end{document}